\documentclass[reqno,11pt]{amsart}
\usepackage{amscd,amssymb,verbatim}
\numberwithin{equation}{section}
\theoremstyle{plain}

\newcommand\alp{\alpha}         
\newcommand\bet{\beta}
\newcommand\gam{\gamma}         \newcommand\Gam{\Gamma}
         
\newcommand\eps{\varepsilon}
\newcommand\zet{\zeta}
\newcommand\tet{\theta}         
\newcommand\iot{\iota}

\newcommand\lam{\lambda}                \newcommand\Lam{\Lambda}
\newcommand\sig{\sigma}         

\newcommand\ome{\omega}         \newcommand\Ome{\Omega}

\newcommand\calI{{\mathcal{I}}}

\newcommand\calS{{\mathcal{S}}}

\newcommand\RR{\mathbb{R}}

\renewcommand\SS{\mathbb{S}}

\newcommand\ZZ{\mathbb{Z}}

\newcommand\CC{\mathbb{C}}

\newcommand\grS{{\mathfrak{S}}}

\newcommand\nek{,\ldots,}
\newcommand\sdp{\times \hskip -0.3em {\raise 0.3ex
\hbox{$\scriptscriptstyle |$}}} % semidirect product

\newcommand\Det{\operatorname{Det}}

\newcommand\dist{\operatorname{dist}}

\newcommand\End{\operatorname{End\,}}

\newcommand\IM{\operatorname{Im}}

\newcommand\Ker{\operatorname{Ker}}

\newcommand\RE{\operatorname{Re}}

\newcommand\spec{{\rm spec\,}}

\newcommand\tr{\operatorname{tr}}
\newcommand\Tr{\operatorname{Tr}}

\newcommand\oD{{\overline{D}}}

\newcommand\on{{\overline{n}}}

\newcommand\hatn{{\widehat{n}}}

\newcommand\tilc{{\widetilde{c}}}

\newcommand\tilD{{\widetilde{D}}}

\renewcommand{\>}{\rangle}
\newcommand{\<}{\langle}

\theoremstyle{plain}
\newtheorem{Thm}[subsection]{Theorem}
\newtheorem{Cor}[subsection]{Corollary}
\newtheorem{Lem}[subsection]{Lemma}
\newtheorem{Prop}[subsection]{Proposition}
\newtheorem{Conjec}[subsection]{Conjecture}

\newtheorem{Def}[subsection]{Definition}%\renewcommand{\theDef}{\thesection.\arabic{Def}}

\theoremstyle{remark}

\newtheorem{Rem}[subsection]{Remark}%\renewcommand{\theRem}{\thesection.\arabic{Rem}}

\def\TeXref#1{%
        \leavevmode\vadjust{\setbox0=\hbox{{\tt
                \  {\tiny \textrm #1}}}%
        \theight=\ht0
        \advance\theight by \lineskip
        \kern -\theight \vbox to
        \theight{\rightline{\rlap{\box0}}%
        \vss}%
        }}%
\newif\ifShowLabels
\ShowLabelstrue
\newdimen\theight
\def\TeXrefEq#1{%
        \leavevmode\vadjust{\setbox0=\hbox{{\tt
                \  {\tiny \textrm #1}}}%
        \theight=\ht1
        \advance\theight by \lineskip
        \kern -\theight \vbox to
        \theight{\rightline{\rlap{\box0}}%
        \vss}%
        }}%
\newcommand{\refs}[1]{Section ~\ref{S:#1}}
\newcommand{\refss}[1]{Subsection ~\ref{SS:#1}}

\newcommand{\reft}[1]{Theorem ~\ref{T:#1}}
\newcommand{\refl}[1]{Lemma ~\ref{L:#1}}

\newcommand{\refd}[1]{Definition ~\ref{D:#1}}
\newcommand{\refr}[1]{Remark ~\ref{R:#1}}
\newcommand{\refe}[1]{\eqref{E:#1}}

\newenvironment{thm}[1]%
        { \begin{Thm} \label{T:#1}  \ifShowLabels \TeXref{T:#1} \fi }%
        { \end{Thm} }

\renewcommand{\th}[1]{\begin{thm}{#1}  }
\renewcommand{\eth}{\end{thm} }

\newenvironment{lemma}[1]%
        { \begin{Lem} \label{L:#1}  \ifShowLabels \TeXref{L:#1} \fi }%
        { \end{Lem} }

\newcommand{\lem}[1]{\begin{lemma}{#1} }
\newcommand{\elem}{\end{lemma}}

\newenvironment{propos}[1]%
        { \begin{Prop} \label{P:#1}  \ifShowLabels \TeXref{P:#1} \fi }%
        { \end{Prop} }

\newcommand{\prop}[1]{\begin{propos}{#1} }
\newcommand{\eprop}{\end{propos}}

\newenvironment{corol}[1]%
        { \begin{Cor} \label{C:#1}  \ifShowLabels \TeXref{C:#1} \fi }%
        { \end{Cor} }
\newcommand{\cor}[1]{\begin{corol}{#1}  }
\newcommand{\ecor}{\end{corol}}

\newenvironment{conjec}[1]%
        { \begin{Conjec} \label{Conj:#1}  \ifShowLabels \TeXref{C:#1} \fi }%
        { \end{Conjec} }
\newcommand{\conj}[1]{\begin{conjec}{#1}  }
\newcommand{\econj}{\end{conjec}}

\newenvironment{defeni}[1]%
        { \begin{Def} \label{D:#1}  \ifShowLabels \TeXref{D:#1} \fi }%
        { \end{Def} }
\newcommand{\defe}[1]{\begin{defeni}{#1}  }
\newcommand{\edefe}{\end{defeni}}

\newenvironment{remark}[1]%
        { \begin{Rem} \label{R:#1}  \ifShowLabels \TeXref{R:#1} \fi }%
        { \end{Rem} }
\newcommand{\rem}[1]{\begin{remark}{#1}}
\newcommand{\erem}{\end{remark}}

\newcommand{\eq}[1]%
        { \ifShowLabels \TeXrefEq{E:#1} \fi
           \begin{equation} \label{E:#1} }
\newcommand{\eeq}{\end{equation}}

\newcommand{\prf}{ \begin{proof} }
\newcommand{\eprf}{ \end{proof} }
\newcommand{\Label}[1]{\label{#1}  \ifShowLabels \TeXref{#1} \fi }

\ShowLabelsfalse% comment this out if labels should be printed

\newcommand{\n}{\nabla}

\newcommand{\pp}{\text{\( \not\!\partial\)}}
\newcommand{\symb}{\sig_L}

\begin{document}

%-------------------------------------
%------------------------------------
\title[Topological calculation of the phase of the
determinant]{Topological calculation of the phase of the determinant of a  non self-adjoint
elliptic operator}
\author{Alexander G. Abanov} \author{Maxim Braverman}
\address{Department of Physics and Astronomy\\
Stony Brook University\\ Stony Brook, NY 11794\\ USA}

\address{Department of Mathematics\\
        Northeastern University   \\
        Boston, MA 02115 \\
        USA
         }
\email{alexandre.abanov@sunysb.edu}\email{maxim@neu.edu}
\thanks{The first author was partially supported by  the Alfred
        P. Sloan foundation. \\ \indent
The second author was partially supported by the NSF grant DMS-0204421.}
%\subjclass{Primary: 32L20; Secondary: 58G10, 14F17}
%\keywords{Vanishing theorem, Clifford bundle, Dirac operator, Andreotti-Grauert
%theorem, Melin inequality}
%\date{\today}
\begin{abstract}
We study the zeta-regularized determinant of a non self-adjoint elliptic operator on a closed
odd-dimensional manifold. We show that, if the spectrum of the operator is symmetric with respect
to the imaginary axis, then the determinant is real and its sign is determined by the parity of the
number of the eigenvalues of the operator, which lie on the positive part of the imaginary axis. It
follows that, for many geometrically defined operators, the phase of the determinant is a
topological invariant. In numerous examples, coming from geometry and physics, we calculate the
phase of the determinants in purely topological terms. Some of those examples were known in
physical literature, but no mathematically rigorous proofs and no general theory were available
until now.
\end{abstract}
\maketitle \tableofcontents

%-----------------------------------------------------
%-----------------------------------------------------

\section{Introduction}\Label{S:introd}

In the recent years several examples appeared in physical literature when the phase of the
determinant of a geometrically defined non self-adjoint Dirac-type operator is a topological
invariant (see e.g., \cite{Redlich84,Redlich84a,AbanovWieg00,Abanov-Hopf}).  Many of those examples
appear in the study of the non-linear $\sig$-model for Dirac fermions coupled to chiral bosonic
fields \cite{AbanovWieg00,Abanov-Hopf}. The topologically invariant phase is called the {\em
$\tet$-term}. It has a dramatic effect on the dynamics of the Goldstone bosons but also has a great
interest for geometers. Unfortunately, no mathematically rigorous proofs of the topological
invariance of the phase of the determinant were available until now.

This paper is an attempt to better understand the above phenomenon. In particular, we find a large
class of operators whose determinants have a topologically invariant phase. We also develop a
technique for calculation of this phase. In particular, we get a first mathematically rigorous
derivation of several examples which appeared in physical literature. In many cases, we also
improve and generalize those examples.

Our first result is \reft{symmetry} which states that {\em the determinant of an elliptic operator
$D$ with a self-adjoint leading symbol, which acts on an odd-dimensional manifold and whose
spectrum is symmetric with respect to the imaginary axis, is real. Moreover, the sign of this
determinant is equal to $(-1)^{m_+}$, where $m_+$ is the number of the eigenvalues of $D$ (counted
with multiplicities) which lie on the positive part of the imaginary axis}.

Note that this result  is somewhat surprising.
Indeed, if one calculates the determinant of a {\em finite} matrix D
with the spectrum symmetric with respect to an imaginary axis, then
one comes to a different result. E.g., the determinant is not
necessarily real.

Suppose now that we are given a family $D(t)$ of operators as above. Assuming that the eigenvalues
of $D(t)$ depend continuously on $t$ one easily concludes (cf. \reft{stability}) that {\em the sign
of the determinant of $D(t)$ is independent of $t$}. In particular, it follows that, if the
definition of the operator $D$ depends on some geometric data (Riemannian metric on a manifold,
Hermitian metric on a vector bundle, etc.), then (provided the spectrum of $D$ is symmetric) {\em
the sign of the determinant is independent of these data, i.e., is a topological invariant}. We
present numerous examples of this phenomenon. In all those examples we calculate the signs of the
determinants in terms of the standard topological invariants, such as
the Betti numbers or the degree of a map.

The paper is organized as follows:

In \refs{determinants}, we briefly recall the basic facts about the $\zet$-regularized determinants
of elliptic operators.

In \refs{symmetry}, we formulate and prove our main result (\reft{symmetry}) and discuss its main
implications.

In \refs{1ex}, we present the simplest (but still interesting) geometric examples of applications
of \reft{symmetry}.

In \refs{circle}, we consider an operator $D$ on a circle, which
appeared in the study of  a quantum spin in the presence of a planar,
time-dependent magnetic field. This operator depends on a map from a circle to itself. We calculate
the phase of the determinant of $D$ in terms of the winding number of this map.

In \refs{degree}, we extend some of the examples considered by P.~Wiegmann and the first author in
\cite{AbanovWieg00}. The operator in question is a Dirac type operator $D$ on an odd dimensional
manifold $M$, whose potential depends on a section $n$ of the bundle of spheres in $\RR\oplus{}TM$.
In particular, if the manifold $M$ is paralelizable, $n$ is a map from $M$ to a
$\dim{}M$-dimensional sphere. We show that the sign of the determinant of $D$ is equal to
$(-1)^{\deg n}$, where $\deg{n}$ is the topological degree of $n$.

\subsection*{Acknowledgments}
The first author  would like to thank the Theory Institute of Strongly Correlated and Complex
Systems at Brookhaven for hospitality and support.

The second author would like to thank the Max-Planck-Institut f\"ur Mathematik, where most of this
work was completed, for hospitality and providing the excellent working conditions. He would also
like to thank Mikhail Shubin and Raphael Ponge for valuable discussions.

%------------------------------------------------------------
%------------------------------------------------------------

\section{Preliminaries on determinants of elliptic operators}\Label{S:determinants}

Let $E$ be a vector bundle over a smooth compact manifold $M$ and let $D:C^\infty(M,E)\to
C^\infty(M,E)$ be an elliptic differential operator of order $m\ge 1$. Let $\symb(D)$ denote the
leading symbol of $D$.

%-------------------------
\subsection{The choice of an angle}\Label{SS:agmonangle}
Our aim is to define the $\zeta$-function and the determinant of $D$. For this we will need to
define the complex powers of $D$. As usual, to define complex powers we need to choose a {\em
spectral cut} in the complex plane. We will restrict ourselves to the simplest spectral cuts given
by a ray
\eq{Rtet}
    R_\tet \ = \ \big\{\, \rho e^{i\tet}:\, 0\le \rho<\infty\, \big\},
    \qquad 0\le \tet \le 2\pi.
\end{equation}
Consequently, we have to choose an angle $\tet\in [0,2\pi)$.

%--------------
\defe{prangle}
The angle $\tet$ is a {\em principal angle} for an elliptic operator $D$ if
\eq{prangle}\notag
    \spec\big(\, \symb(D)(x,\xi)\, \big)\ \cap\ R_\tet \ = \ \emptyset,
    \qquad \text{for all}\quad x\in M,\ \xi\in T^*_xM\backslash\{0\}.
\end{equation}
\edefe

If $\calI\subset \RR$ we denote by $L_{\calI}$ the solid angle
\eq{Lab}\notag
    L_{\calI} \ = \  \big\{\, \rho e^{i\tet}:\, 0 < \rho<\infty,\, \tet\in \calI\,  \big\}.
\end{equation}

\defe{Agmon}
The angle $\tet$ is an {\em Agmon angle} for an elliptic operator $D$ if it is principal angle for
$D$ and there exists $\eps>0$ such that
\[
    \spec(D) \, \cap L_{[\tet-\eps,\tet+\eps]} \ = \ \emptyset.
\]
\edefe

%-------------------------
\subsection{The $\zet$-function and the determinant}\Label{SS:zet-det}
Let $\tet$ be an Agmon angle for $D$. Assume, in addition, that $D$ is injective. The
$\zeta$-function $\zeta_\tet(s,D)$ of $D$ is defined as follows.

Let $\rho_0>0$ be a small number such that
\eq{rho0}\notag
    \spec(D) \, \cap \, \big\{\, z\in \CC;\, |z|<2\rho_0\, \big\} \ = \ \emptyset.
\end{equation}
Define the contour $\Gam= \Gam_{\tet,\rho_0}\subset \CC$ consisting of three curves $\Gam=
\Gam_1\cup \Gam_2\cup \Gam_3$, where
\begin{gather}\Label{E:Gamtetrho}\notag
    \Gam_1 \ = \ \big\{\, \rho e^{i\tet}:\, \rho_0\le \rho<\infty\, \big\},
    \quad
    \Gam_2 \ = \ \big\{\, \rho_0 e^{i\alp}:\, \tet< \alp<\tet+2\pi\, \big\},\\
    \quad
    \Gam_3 \ = \ \big\{\, \rho e^{i(\tet+2\pi)}:\, \rho_0\le \rho<\infty\, \big\}.
\end{gather}

Assume that $\tet\not=0$. For $\RE s> \frac{\dim M}m$, the operator
\eq{Ds}
    D_\tet^{-s} \ = \ \frac{i}{2\pi}\, \int_{\Gam_{\tet,\rho_0}}\, \lam^{-s}_\tet(D-\lam)^{-1}\, d\lam
\end{equation}
is a pseudo-differential operator with smooth kernel $D_\tet^{-s}(x,y)$, cf. \cite{Seeley67, ShubinPDObook}. Here $\lam^{-s}_\tet:=
e^{-s\log_\tet \lam}$ where $\log_\tet\lam$ denotes the branch of the logarithm in $\CC\backslash{R_\tet}$ which takes real values on the
positive real axis.

We define
\eq{zeta}
    \zeta_{\tet}(s,D) \ = \ \Tr D_\tet^{-s} \ = \ \int_M\, \tr D_\tet^{-s}(x,x)\, dx,
    \qquad \RE s> \frac{\dim M}m.
\end{equation}
It was shown by Seeley \cite{Seeley67} (see also \cite{ShubinPDObook}) that $\zeta_{\tet}(s,D)$ has
a meromorphic extension to the whole complex plane and that 0 is a regular value of
$\zeta_{\tet}(s,D)$.

More generally, let $Q$ be a pseudo-differential operator of order $q$. We set
\eq{zetaQ}
    \zeta_{\tet}(s,Q,D) \ = \ \Tr\, Q\,D_\tet^{-s},
    \qquad \RE s> (q+\dim M)/m.
\end{equation}
If $Q$ is a projection, i.e., $Q^2=Q$ then \cite[\S6]{Wodzicki84}, \cite{Wodzicki87} (see
also \cite{Ponge-asymetry} for a shorter proof), the function
$\zeta_{\tet}(s,D;Q)$ also has a meromorphic extension to the whole
complex plane which is regular at 0.

Finally, we define the $\zeta$-regularized determinant of $D$ by the formula
\eq{zeta-det}
    \Det_\tet(D) \ = \ \exp\, \Big(\, -\frac{d}{ds}\big|_{s=0}\zeta_{\tet}(s,D)\, \Big).
\end{equation}
%We will also use the notation
%\eq{logdet}
%    \log\Det_\tet D \ := \ -\frac{d}{ds}\big|_{s=0}\zeta_{\tet}(s,D).
%\end{equation}

%-------------------------------------------
\subsection{The case of an operator close to self-adjoint}\Label{SS:det-sa}
Let us assume now that
\eq{close2sa}
    \symb(D)^*  \ = \ \symb(D),
\end{equation}
where $\symb(D)^*$ denotes the dual of $\symb(D)$ with respect to some fixed scalar product on the
fibers on $E$. This assumption implies that $D$ can be written as a sum $D=D'+A$ where $D'$ is
self-adjoint and $A$ is a differential operator of a smaller order. In this situation we say that
$D$ is {\em an operator close to self-adjoint}, cf. \cite[\S6.2]{Agranovich90},
\cite[\S{}I.10]{Markus88}.

Though the operator $D$ is not self-adjoint in general, the assumption \refe{close2sa} guarantees that it has nice spectral properties. More
precisely, cf. \cite[\S{}I.6]{Markus88}, the space $L^2(M,E)$ of square integrable sections of $E$  is the closure of the algebraic direct sum
of finite dimensional $D$-invariant subspaces
\eq{L=sumLam}
    L^2(M,E) \ = \ \overline{\bigoplus \Lam_k}
\end{equation}
such that the restriction of $D$ to $\Lam_k$ has a unique eigenvalue $\lam_k$ and $\lim_{k\to\infty}|\lam_k|=\infty$. In general, the sum
\refe{L=sumLam} is not a sum of mutually orthogonal subspaces.

The space $\Lam_k$ are called the {\em space of root vectors of $D$ with eigenvalue $\lam_k$}. We
call the dimension of the space $\Lam_k$  the {\em multiplicity} of the eigenvalue $\lam_k$ and we
denote it by $m_k$.

By Lidskii's theorem \cite{Lidskii59}, \cite[Ch.~XI]{Retherford93}, the $\zet$-function \refe{zeta} is equal to the sum (including the
multiplicities)  of the eigenvalues of $D_\tet^{-s}$. Hence,
\eq{zeta-lam}
    \zet_{\tet}(s,D) \ = \ \sum_{k=1}^\infty\, m_k\lam_k^{-s}
    \ = \ \sum_{k=1}^\infty\, m_k e^{-s \log_\tet \lam_k},
\end{equation}
where $\log_\tet(\lam_k)$ denotes the branch of the logarithm in $\CC\backslash{R_\tet}$ which take
the real values on the positive real axis.

%-------------------------------------
\subsection{Dependence of the determinant on the angle}\Label{SS:det-tet}
Assume now that  $\tet$ is only a principal angle for $D$. Then, cf. \cite{Seeley67,
ShubinPDObook}, there exists $\eps>0$ such that $\spec(D)\cap L_{[\tet-\eps,\tet+\eps]}$ is finite
and $\spec(\symb(D))\cap L_{[\tet-\eps,\tet+\eps]}=\emptyset$. Thus we can choose an Agmon angle
$\tet'\in (\tet-\eps,\tet+\eps)$ for $D$. In this subsection we show that $\Det_{\tet'}(D)$ is
independent of the choice of this angle $\tet'$. For simplicity, we will restrict ourselves with
the case when $D$ is an operator close to self-adjoint, cf. \refss{det-sa}.

Let $\tet''> \tet'$  be another Agmon angle for $D$ in $(\tet-\eps,\tet+\eps)$. Then there are only
finitely many eigenvalues $\lam_{r_1}\nek \lam_{r_k}$ of $D$ in the solid angle
$L_{[\tet',\tet'']}$. We have
\eq{tet-tet}
    \log_{\tet''}\lam_k \ = \
    \begin{cases}
        \log_{\tet'}\lam_k, \quad&\text{if}\ \ k\not\in \{r_1\nek r_k\};\\
        \log_{\tet'}\lam_k+2\pi i, \quad&\text{if}\ \ k\in \{r_1\nek r_k\}.
    \end{cases}
\end{equation}
Hence,
\eq{zet-zet}
    \zet_{\tet'}'(0,D) \ - \ \zet_{\tet''}'(0,D) \ = \
    \frac{d}{ds}\big|_{s=0}\,
     \sum_{i=1}^k\, m_k\,  e^{-s\log_{\tet'}(\lam_{r_i})}(1-e^{-2\pi i s})
    \ = \ 2\pi i\, \sum_{i=1}^k\, m_{r_i}.
\end{equation}
%
%In other words, we have
%\eq{logdet-logdet}
%    \log\Det_{\tet''} D \ = \ \log\Det_{\tet'}-2\pi i\, \sum_{i=1}^k\, m_{r_i},
%\end{equation}
and
\eq{det-det}
        \Det_{\tet''} D \ = \ \Det_{\tet'}D.
\end{equation}

Note that the equality \refe{det-det} holds only because both angels $\tet'$ and $\tet''$ are close
to a given principal angle $\tet$ so that the intersection \/ $\spec(D)\cap L_{[\tet',\tet'']}$\,
is finite. If there are infinitely many eigenvalues of $D$ in the solid angle $L_{[\tet',\tet'']}$
then $\Det_{\tet''}(D)$ and $\Det_{\tet'}(D)$ might be quite different.

%---------------------------------------------------------------
%---------------------------------------------------------------
\section{Operators whose spectrum is symmetric with respect to the imaginary axis}\Label{S:symmetry}

In this section $M$ is an {\em odd-dimensional} closed manifold, $E\to M$ is a complex vector
bundle over $M$, and $D$ is a differential operator of order $m\ge1$ which is close to self-adjoint
(cf. \refss{det-sa}) and invertible.

%------------------
\subsection{The phase of the determinant and the imaginary eigenvalues}\Label{SS:nofeig}
Suppose that {\em the spectrum of $D$ is symmetric with respect to the imaginary axis}. More precisely, we assume that, if $\lam=
\rho{}e^{i\alp}$ is an eigenvalue of $D$ with multiplicity $m$, then $\rho{}e^{-i(\pi+\alp)}$ is also an eigenvalue of $D$ with the same
multiplicity. Since the leading symbol of $D$ is self-adjoint, $\pm\frac\pi2$ are principal angles of $D$, cf. \refd{prangle}. Hence, cf.
\refss{det-tet}, we can choose an Agmon angle $\tet\in (\frac\pi2,\pi)$ such that there are no eigenvalues of $D$ in the solid angles
$L_{(\pi/2,\tet]}$ and $L_{(-\pi/2,\tet-\pi]}$.

Let $m_+$ denote the number of eigenvalues of $D$ (counted with multiplicities) on the positive
part of the imaginary axis, i.e., on the ray $R_{\pi/2}$ (cf. \refe{Rtet}).

Our first result is the following
\th{symmetry}
In the situation described above
\eq{symmetry}
    \IM \zet_\tet'(0,D) \ = \ -\, \pi\, m_+.
\end{equation}
In particular, $\Det_\tet D = \exp\big(\,-\zet_\tet'(0,D)\, \big)$ is a real number, whose sign is
equal to $(-1)^{m_+}$.
\eth
\rem{depontet}
a. \ For \refe{symmetry} to hold we need the precise assumption on $\tet$ which we specified above.
However, if we are only interested in the sign of the determinant of $D$, the result remains true
for all $\tet\in (-\pi,\pi)$. This follows from \refe{det-det}.

b. \ Note that only the eigenvalues on the positive part of the imaginary axis contribute to the
sign of the determinant. This asymmetry between the positive and the negative part of the imaginary
axis is coursed by our choice of the spectral cut $R_\tet$ in the upper half plane. If we have
chosen the spectral cut in the lower half plane the sign of the determinant would be determined by
the eigenvalues on the negative imaginary axis.
\erem
%%%
%--------------------------
\prf
Let
\[
    \rho_je^{i\alp_j}, \qquad \tet-\pi<\alp_j<\frac\pi2, \quad j=1,2,\ldots
\]
be all the eigenvalues of $D$ which lie in the solid angle $L_{(\tet-\pi,\pi/2)}$ (here and below all the eigenvalues appear in the list the
number of times equal to their multiplicities). Since the spectrum of $D$ is symmetric with respect to the imaginary axis,
$\rho_je^{-i(\pi+\alp_j)}$ $(j=1,2,\ldots)$ are all the eigenvalues of $D$ in the solid angle $L_{(\tet-2\pi,-\pi/2)}$.%
\footnote{Since we have chosen a spectral cut along the ray $R_\tet$ with $\tet\in (\pi/2,\pi)$ we write all the eigenvalues in the form
$\lam=\rho{}e^{i\alp}$ with $\alp\in (\tet-2\pi,\tet)$ so that $\lam_\tet^{-s}= \rho^{-s}e^{-i\alp s}$.}

Finally, let
\[
    \mu_1^+\,e^{i\frac\pi2}\nek \mu_{m_+}^+\, e^{i\frac\pi2}, \qquad
    \mu_1^-\,e^{-i\frac\pi2}\nek \mu_{m_-}^-\, e^{-i\frac\pi2}
\]
be all the imaginary eigenvalues of $D$ (since $\pm\frac\pi2$ are principal angles for $D$, there
are only finitely many of those, cf. \refss{det-tet}). Then
\begin{multline}\Label{E:zetsD}\notag
    \zet_\tet(s,D) \ = \
    \sum_{j=1}^\infty\, \rho_j^{-s}\, \big(\, e^{-i\alp_j s}+ e^{i(\alp_j+\pi)s}\,\big)
    \ + \
    \sum_{j=1}^{m_+}\, (\mu_j^+)^{-s}\, e^{-i\frac\pi2 s}
    \ + \ \sum_{j=1}^{m_-}\, (\mu_j^-)^{-s}\, e^{i\frac\pi2 s}
    \\ = \
    2\,\Big[\,
     \sum_{j=1}^\infty\, \rho_j^{-s}\, \cos\,(\alp_j+\frac\pi2)s\, \Big]\,e^{i\frac\pi2s}
    \ +\ \sum_{j=1}^{m_+}\, (\mu_j^+)^{-s}\, e^{-i\frac\pi2 s}
    \ + \ \sum_{j=1}^{m_-}\, (\mu_j^-)^{-s}\, e^{i\frac\pi2 s}.
\end{multline}
Set
\[
    z(s) \ := \ 2\, \sum_{j=1}^\infty\, \rho_j^{-s}\, \cos\,(\alp_j+\frac\pi2)s.
\]
Then
\[
    \zet_\tet(s,D) \ = \ z(s)\,e^{i\frac\pi2s} \ + \
    \sum_{j=1}^{m_+}\, (\mu_j^+)^{-s}\, e^{-i\frac\pi2 s}
    \ + \ \sum_{j=1}^{m_-}\, (\mu_j^-)^{-s}\, e^{i\frac\pi2 s},
\]
and
\eq{zet'0D}\notag
    \zet'_\tet(0,D) \ = \ z'(0) \ + \ i\frac\pi2\, z(0)\ + \
    \sum\, \log\mu_j^\pm \ + \ i\frac\pi2\, \big(\, m_--m_+\,\big).
\end{equation}
Note that $z(s)$ and $z'(s)$ are real for $s\in \RR$. Hence, we obtain
\eq{imzet'0}
    \IM\, \zet'_\tet(0,D) \ = \
    \frac\pi2\, \big(\, z(0) + m_- - m_+\, \big).
\end{equation}

We will now calculate $z(0)$ by comparing it with the $\zet$-function of the operator $D^2$. The
angle $2\tet$ is a principal angle for $D^2$ and
\begin{multline}\notag
    \zet_{2\tet}(s/2,D^2) \ = \
    \sum\, \rho_j^{-s}\, \big(\, e^{-i\alp_j s}+ e^{i\alp_js}\,\big)
    \ + \
    \sum\, (\mu_j^\pm)^{-s}\, e^{-i\frac\pi2 s}
    \\ = \ 2\, \sum\, \rho_j^{-s}\, \cos(\alp_js)
            \ + \ \sum\, (\mu_j^\pm)^{-s}\, e^{-i\frac\pi2 s}.
\end{multline}
Hence,
\eq{zet-z}
    \zet_{2\tet}(s/2,D^2) \ - \  z(s) \ = \
    4\, \Big[\,\sum\, \rho_j^{-s}\,\sin(\alp_j+\frac\pi4)s\, \Big]\, \sin\frac\pi4s
    \ + \  \sum\, (\mu_j^\pm)^{-s}\, e^{-i\frac\pi2 s}.
\end{equation}

Let $\Pi_{(-\pi/2,\pi/2)},\, \Pi_{(\pi/2,3\pi/2)}:L^2(M,E)\to L^2(M,E)$ be the orthogonal
projections onto the spans of the eigensections of $D$ corresponding to the eigenvalues in
$L_{(-\pi/2,\pi/2)}$ and in $L_{(\pi/2,3\pi/2)}$ respectively. Then, using the notation introduces
in \refe{zetaQ}, we obtain
\begin{multline}\notag
    \zet_{\tet}(s,\Pi_{(-\pi/2,\pi/2)},D)\, e^{-i\frac\pi4 s}
    \ - \ \zet_{\tet}(s,\Pi_{(\pi/2,3\pi/2)},D)\, e^{-i\frac{3\pi}4 s}
    \\ =\
    \Big[\, \sum\, \rho_j^{-s}\,
                    e^{-i\alp_js}\,\Big]\,e^{-i\frac{\pi}4s}
    \ -\
    \Big[\, \sum\, \rho_j^{-s}\,
                    e^{i(\alp_j+\pi)s}\,\Big]\,e^{-i\frac{3\pi}4s}
    \\ = \
    \sum\, \rho_j^{-s}\,\, \Big(\, e^{-i(\alp_j+\frac\pi4)}-e^{i(\alp_j+\frac\pi4)}\,\Big)
    \ = \
    -2i\, \sum\, \rho_j^{-s}\,\sin(\alp_j+\frac\pi4)s.
\end{multline}
Hence, cf. the discussion in the end of \refss{zet-det}, the function $\sum\,
\rho_j^{-s}\,\sin(\alp_j+\frac\pi4)s$ has a meromorphic extension to the whole complex plane, which
is regular at 0. Thus, the first term in the RHS of \refe{zet-z} vanishes when $s=0$. The equality
\refe{zet-z} implies now that
\[
    \zet_{2\tet}(0,D^2) \ - \ z(0) \ = \ m_+ + m_-.
\]
It is well known, cf. \cite{Seeley67}, that the $\zet$-function of a differential operator of even
order on an odd-dimensional manifold vanishes at 0. In particular, $\zet_{2\tet}(0,D^2)=0$. Thus,
\[
    z(0) \ = \ -\,\big(\, m_+ + m_-\,\big).
\]
Substituting this equality into \refe{imzet'0}, we obtain \refe{symmetry}.
\eprf

%------------
\rem{surprise}
Note that the result of \reft{symmetry} is somewhat surprising. Indeed, if one thinks about
$\Det_\tet{D}$ as a formal product of the eigenvalues of $D$, then one can do the following formal
computation (where we shall use the notation introduces in the proof of \reft{symmetry}): for each
$j=1,2,\ldots$ the product of the eigenvalues $\rho_j{}e^{i\alp_j}$ and $\rho_j{}e^{i(\pi-\alp_j)}$
is a real number. Hence, one expects $\Det_\tet{D}=
\pm\big|\Det_\tet{D}\big|e^{i\frac\pi2(m_+-m_-)}$, which is quite different from the correct answer
given by \reft{symmetry}.

This example illustrates the danger of formal manipulations with determinants%
\footnote{The fact that formal computations often lead to wrong answers is well known. In
particular, $\Det_\tet{D}$ might not be real even if $D=D^*$ so that all the eigenvalues of $D$ are
real, cf., for example, \cite{Wojciechowski99}.}.
\erem

%--------------------------------------------------------------------------
\subsection{Stability of the phase of the determinant}\Label{SS:stqability}
Suppose now that $D(t)$ is a family of close to self-adjoint operators, depending on a real parameter $t$. We will say that {\em the spectrum of
$D(t)$ depends continuously on $t$} if, for each $t$, we can represent $L^2(M,E)$ as a closure of a sum of $D(t)$-invariant finite dimensional
subspaces,
\[
    L^2(M,E) \ = \ \overline{\bigoplus\, \Lam_k(t)},
\]
(cf. \refe{L=sumLam}) such that
\begin{itemize}
\item
$\dim\Lam_k(t)$ is independent of $t$;
\item
the restriction of $D(t)$ to $\Lam_k$ has a unique eigenvalue $\lam_k(t)$ and
$\lim_{k\to\infty}|\lam_k(t)| =0$;
\item
for every $k=1,2,\ldots$, the function $\lam_k(t)$ is continuous in $t$.
\end{itemize}

%------
\th{stability}
Let now $D(t)$ be a family of operators depending on a real parameter $t$. We assume that for each
$t\in \RR$ the operator $D(t)$ satisfies all the assumptions of \reft{symmetry}. In particular, its
spectrum is symmetric with respect to the imaginary axis. Assume, in addition, that the eigenvalues
of $D(t)$ depend continuously on $t$. For each $t$ let us choose an Agmon angle $\tet(t)\in
(\pi/2,\pi)$. Then  $\Det_{\tet(t)}D(t)$ is real and its sign of is independent of $t$.
\eth
\prf
By our assumptions, the eigenvalues of $D(t)$ are symmetric with respect to the imaginary axis and
never pass through zero.  It follows that when one of the eigenvalues reaches $R_{\pi/2}$ from the
left the other must reach it from the right. In other words, the parity of the number of the
eigenvalues on the imaginary axis is independent of $t$. The theorem follows now from
\reft{symmetry} and \refr{depontet}.
\eprf

%---------------------------------------------------------------------------
\subsection{Topological invariance of the phase of the determinant}\Label{SS:topofphase}
The eigenvalues always depend continuously on $t$ if $D(t)=D_0+tB$, where $D_0$ is an elliptic
differential operator of order $m\ge1$ and $B$ is a differential operator whose order is less than
$m$, cf. \cite{Kato}. They also often depend continuously on $t$ when we have a smooth family of
geometric structures (i.e, Riemannian metrics on a manifold, Hermitian metrics on a vector bundle,
etc.) and $D(t)$ is a family of geometrically defined operators (Dirac operators, Laplacians, etc.)
depending of these geometric structures. Suppose, in addition, the spectrum of $D(t)$ is symmetric
with respect to the imaginary axis. Then, in view of \reft{stability},
it is natural to expect that the
phase of the determinant is {\em a topological invariant}. A natural question is how to relate this
invariant to the other topological invariants. In other words, we would like to find a topological
method of calculating the phase of the determinant of geometric operators, whose spectrum is
symmetric with respect to the imaginary axis. In the rest of the paper we present numerous examples
in which such a calculation is indeed possible.

%-------------------------------------------------------------
%-------------------------------------------------------------

\section{First examples}\Label{S:1ex}

In this section we present some simple examples of applications of \reft{symmetry}. More
sophisticated examples will be considered in the subsequent sections.

%-----------------------------------------------
\subsection{The circle}\Label{SS:circle1}
The simplest possible example of an operator satisfying the conditions of \reft{symmetry} is the
operator
\eq{circle1}\notag
    D_a \ = \ -i\,\frac{d}{dt} \ + \ ia, \qquad a\in \RR,
\end{equation}
acting on the space of function on the circle $S^1$. Clearly, the only imaginary eigenvalue of
$D_a$ is $ia$. Hence, \reft{symmetry} implies that, for $\tet\in (0,\pi)$, we have
\eq{Detcircle1}
    \Det_\tet D_a \ < \ 0, \ \ \text{if} \ \ a>0, \quad \text{and}
    \quad \Det_\tet D_a \ > \ 0  \ \text{if} \ \ a<0.
\end{equation}

%-------------
\rem{circle}
The determinant of the operator $D_a$ can be calculated explicitly. In fact \cite{BFK91} provides a
formula for this determinant in terms of the monodromy operator associated to $D_a$. Using this
formula, one easily gets
\[
   \Det_\tet D_a \ = \ e^{-a\bet}-1,
\]
where $\bet$ is the length of the circle. Clearly, that agrees with \refe{Detcircle1}.
\erem

%--------------------------------------------------------
\subsection{The deformed DeRham-Dirac operator}\Label{SS:deRhamdirac}
Suppose $M$ is a closed manifold of odd dimension $N=2l+1$. Let $d:\Ome^*(M)\to \Ome^{*+1}(M)$ be
the DeRham differential and let $d^*:\Ome^*(M)\to \Ome^{*-1}(M)$ be the adjoint of $d$ with respect
to a fixed Riemannian metric on $M$. Let $\bet_j(M)= \dim{}H^j(M)$ ($j=0\nek N$) denote the Betti
numbers of $M$.

The operator
\[
   D_a \ := \ d \ +\ d^*\ + \ ia, \qquad a\in \RR, \ \ a\not=0.
\]
has exactly one imaginary eigenvalue $\lam= ia$ and its multiplicity is equal to the sum
$\sum_{j=0}^N\bet_j(M)$ of the Betti numbers of $M$. Because of the Poincar\'e duality, this sum is
an even number. Thus \reft{symmetry} implies that
\[
   \Det_\tet\, D_a \ > \ 0, \qquad \text{for all} \quad a\in \RR, \ 0<\tet<\pi.
\]

To construct a more interesting example let us fix non-zero real numbers $a_0\nek a_N$ and consider
the operator $A:\Ome^*(M)\to \Ome^*(M)$ defined by the formula
\[
   A:\, \ome\ \mapsto a_j\,\ome, \qquad \text{if}\quad \ome\in
   \Ome^j(M).
\]
Then one easily concludes from \reft{symmetry} that
\[
   \Det_\tet\, \big(\, d+d^*+iA\, \big) \ = \ (-1)^{\sum_{\{j:a_j>0\}}\bet_j(M)}\,
   \big|\, \Det_\tet\, \big(\, d+d^*+iA\, \big)\, \big|.
\]

Another interesting example can be constructed as follows. Let $*:\Ome^*(M)\to \Ome^{N-*}(M)$
denote the Hodge-star operator. Consider the operator $\Gam:\Ome^*(M)\to \Ome^{N-*}(M)$, defined by
the formula
\eq{chiral}
    \Gam:\, \alp \ \mapsto \
    i^{\frac{N(N+1)}2}\,(-1)^{\frac{j(j+1)}2}\, *\alp
    \ = \
    i^{l+1}\,(-1)^{\frac{j(j+1)}2}\, *\alp,
    \qquad \alp\in \Ome^j(M).
\end{equation}
Since $N=2l+1$ is odd, $\Gam$ is self-adjoint, satisfies $\Gam^2=1$, and commutes with $d+d^*$. In
particular, $\Gam$ acts on $\Ker(d+d^*)$ and this action has exactly 2 eigenvalues $\pm1$, which
have equal multiplicities $\frac12\sum_{j=0}^N\bet_j(M)$. Hence, the operator
\[
   D_\Gam \ := \ d\ +\ d^*\ + \ i\,\Gam,
\]
has exactly 2 imaginary eigenvalues $\pm{i}$, and multiplicities of these eigenvalues are equal to
$\frac12\sum_{j=0}^N\bet_j(M)$. \reft{symmetry} implies now that
\[
   \Det_\tet\, D_\Gam \ = \ (-1)^{\frac12\sum_{j=0}^N\bet_j(M)}\,
   \big|\, \Det_\tet\, D_\Gam\, \big|.
\]

%------------------
\rem{d-nabla}
All the results of this subsection can be easily extended to operators acting on the space of
differential forms with values in a flat vector bundle $F\to M$. (The DeRham differential should be
replaced by the covariant differential and the Betti numbers should be replaced by the dimensions
of the cohomology of $M$ with coefficients in $F$). We leave the details to the interested reader.
\erem

%----------------------------------------
%------------------------------------------
\section{A Dirac-type operator on a circle}\Label{S:circle}

In this section we consider the operator $D$ on the circle, which appears, e.g., in the study of  a quantum spin in the presence of a planar,
time-dependent magnetic field. In the case when magnetic field is changing adiabatically in time the wave function of spin 1/2 acquires the
phase $\pi k$, where $k$ is an integer number of rotations that the direction $n$ of magnetic field makes around the origin during the time
evolution. This adiabatic phase is called the Berry phase \cite{Berry84}. For adiabatic evolution of magnetic field the Berry phase is equal up
to a trivial dynamic factor to the determinant of the operator $D$ defined in \refe{D} below, cf., e.g., \cite{AbanovWieg00}. The main result of
this section is \reft{circle} which calculates the phase of this determinant. This theorem is known in physical literature (see e.g.,
\cite{AbanovWieg00}), but no mathematically rigorous proofs were available until now.

%-----------------
\subsection{The setting}\Label{SS:setting-circle}
Let $S^1$ be the circle, which we view as the interval $[0,\bet]$
($\bet>0$) with identified ends. Let
\[
   n:\, S^1 \ \longrightarrow \ \big\{\, z\in \CC:\, |z|=1\, \big\}
\]
be a smooth map. Then there exists a smooth function $\phi:\RR\to \RR$,
satisfying the periodicity conditions
\eq{phi=phi+k}
    \phi(t+\bet) \ = \ 2\pi k+ \phi(t), \qquad k\in \ZZ,
\end{equation}
such that $n=e^{i\phi}$. The number $k$ above is called the {\em
  topological degree} (or the {\em winding number}) of the map $n$.

Set $\hatn=\Big(\begin{smallmatrix}0&e^{i\phi}\\
e^{-i\phi}&0\end{smallmatrix}\Big)$ and consider the family of
operators depending on a real parameter $m$
\eq{D}
    D \ =  \ i\frac{d}{dt}+im\hatn
       \ = \ i \frac{d}{dt}+ i m \begin{pmatrix}0&e^{i\phi}\\ e^{-i\phi}&0\end{pmatrix},
\end{equation}
acting on the space of vector-functions $\xi:[0,\bet]\to \CC^2$ with boundary conditions
\eq{boundary}
    \xi(\bet)=e^{i\pi\nu}\xi(0), \quad \dot{\xi}(\bet)=e^{i\pi\nu}\dot{\xi}(0), \quad
    \qquad \nu=0,1.
\end{equation}

We shall study the determinant of $D$. The following lemma shows that this determinant is non-zero
for $m$ sufficiently large.

%------------
\lem{nokernel}
For $m> \max_{t\in[0,\bet]}|\dot{\phi}(t)|$, zero is not in the spectrum of $D$.
\elem
%%%
%%%
\prf
Consider the following scalar product on the vector valued functions on $[0,\bet]$:
\[
    (\xi, \eta) \ = \ \int_0^\bet\, \<\xi(t),\eta(t)\>\, dt,
\]
where $\<\cdot,\cdot\>$ stands for the standard scalar product on $\CC^2$.  Let
$\|\xi\|=(\xi,\xi)^{1/2}$ denote the norm of the vector function $\xi$.

Integrating by parts the expression for $\|D\xi\|^2$ we obtain, for  $\xi$ satisfying the boundary
condition \refe{boundary},
\begin{multline}\notag
    \|D\xi\|^2 \ = \ (\dot\xi,\dot\xi)+m(\hatn\xi,\dot\xi)+m(\dot\xi,\hatn\xi)+m^2\|\xi\|^2
    \\ \ge \ -m(\dot{\hatn}\xi,\xi)-m(\hatn\dot\xi,\xi)+m(\dot\xi,\hatn\xi)+m^2\|\xi\|^2 \\
    = \ -m(\dot\hatn\xi,\xi)+m^2\|\xi\|^2
     \ \ge \ m\, \big(\, m-\max_{t\in[0,\bet]}\, |\dot\phi(t)|\, \big)\, \|\xi\|^2.
\end{multline}
\eprf

%---------
\th{circle}
Let $m>\max_{t\in [0,\bet]}|\dot{\phi}(t)|$. For every $\tet\in (0,\pi)$ such that there are no
eigenvalues of $D$ on the ray $R_\tet$ the following equality holds
\eq{main}
    \Det_\tet D \ = \ -(-1)^{k+\nu}\, \big|\,\Det_\tet D\, \big|,
\end{equation}
where $k$ is defined in \refe{phi=phi+k} and $\nu$ is defined in \refe{boundary}.
\eth
\rem{top-circle}
\reft{circle} relates the sign of $\Det_\tet{D}$ with
the topological invariant of the map $e^{i\phi}$. This realizes the program outlined in
\refss{topofphase}.
\erem

We precede the proof of the theorem with some discussion of the spectral properties of $D$.

%----------------------------------
\subsection{The spectral properties of $D$}\Label{SS:spectrum}
In order to study the spectrum of $D$ it is convenient to replace it by a conjugate operator as
follows. The operator
\[
    U_\phi \ := \   \begin{pmatrix}
                        e^{i\phi/2}&0\\ 0&e^{-i\phi/2}
                    \end{pmatrix}
\]
maps the space of vector-functions with boundary conditions \refe{boundary} to the space of
functions $\xi:[0,\bet]\to \CC^2$ with new boundary conditions
\eq{boundary2}
    \xi(\bet)=e^{i\pi(\nu+k)}\xi(0), \quad \dot{\xi}(\bet)=e^{i\pi(\nu+k)}\dot{\xi}(0).
\end{equation}
Thus the operator
\eq{tilD}
    \tilD \ := \ U_\phi^{-1}\circ D\circ U_\phi \ = \
    i\, \frac{d}{dt}+  \begin{pmatrix}-\dot{\phi}/2&im\\ im&\dot{\phi}/2\end{pmatrix}
\end{equation}
acting on the space of vector functions with boundary conditions \refe{boundary2} is isospectral to
$D$.

We now consider the following deformation of $\tilD$:
\eq{tilDa}
    \tilD_a \ := \
    i\, \frac{d}{dt}+
      \begin{pmatrix}-a\dot{\phi}/2&im\\ im&a\dot{\phi}/2\end{pmatrix},
    \qquad a\in [0,1].
\end{equation}
The same arguments which were used in the proof of \refl{nokernel} show that

\lem{nokernel2}
The operator $\tilD_a$ is invertible for all $a\in [0,1]$, and all sufficiently large $m>0$.
\elem

%--------------------------
Let
\[
    \overline{\tilD_a} \ = \    - i\, \frac{d}{dt}+
      \begin{pmatrix}-a\dot{\phi}/2&-im\\ -im&a\dot{\phi}/2\end{pmatrix},
\]
be the complex conjugate of the operator $\tilD_a$.

The following lemma shows that the spectrum of $\tilD_a$ (and, hence, of $D$) is symmetric with
respect to both the real and the imaginary axis.
%---------
\lem{symmetry}
The operators $\tilD_a$, $-\overline{\tilD}_a$, and $\tilD^*_a$ are conjugated to each other.
Therefore, they have the same spectral decomposition \refe{L=sumLam}.

In particular, the operators $D$, $-\oD$, and $D^*$ are conjugated to each other.
\elem
%------
\prf
An easy calculation shows that
\[
    \begin{pmatrix} 1&0\\0&-1\end{pmatrix}\cdot \tilD_a \cdot \begin{pmatrix} 1&0\\0&-1\end{pmatrix}
    \ = \ \tilD_a^*, \qquad
    \begin{pmatrix} 0&1\\1&0\end{pmatrix}\cdot
      \tilD_a \cdot\begin{pmatrix}
    0&1\\1&0\end{pmatrix} \ = \ -\overline{\tilD}_a.
\]
\eprf

%--------
\subsection{Proof of \reft{circle}}\Label{SS:prcircle}
Since the operators $D$ and $\tilD$ are conjugated to each other their determinants are equal. By
\reft{stability}, the sign of the determinant of $\tilD=\tilD_a$ is equal to the sign of the
determinant of the operator
\[
    \tilD_0 \ = \     i\, \frac{d}{dt}+
      \begin{pmatrix}0&im\\ im&0\end{pmatrix}.
\]
It is easy to see that all the eigenvalues of $\tilD_0$ are given by the formula
\eq{sptilD0}\notag
    \lam^\pm_n \ = \ \pm im \ + \ \frac{\pi}{\bet}\,(2n-k-\nu), \qquad n\in \ZZ.
\end{equation}
Hence, $\tilD_0$ does not have any eigenvalues on the ray $R_{\pi/2}$ if $k+\nu$ is odd and has
exactly one eigenvalue $\lam^+_{(k+\nu)/2}= im$ on this ray if $k+\nu$ is even.

\reft{circle} follows now from \reft{symmetry}. \hfill$\square$

%--------------------------------------------------------
%--------------------------------------------------------
\section{The phase of the determinant and the degree of the
  map}\Label{S:degree}

This section essentially generalizes the previous section to manifolds of higher dimensions. For
the case of a sphere of dimension $N=4l+1$ the results of this section have been obtained in
\cite{AbanovWieg00} as topological terms in non-linear $\sig$-models emerging as effective models
for Dirac fermions coupled to chiral bosonic fields. However, no mathematically rigorous proofs
were available until now. Note also that our result is more precise, since the equality
\refe{degree} was obtained in \cite{AbanovWieg00}  from the  gradient expansion, i.e.,  only
asymptotically for $m\to\infty$.

The section is organized as follows: first we formulate the problem in purely geometric terms as a
question about the determinant of the DeRham-Dirac operator with potential. We state our main
result as \reft{degree}. Then, in \refss{reform}, we reformulate the result in terms of an operator
acting on the tensor product of the two spaces of spinors. This formulation is closer to the one
considered in physical literature. Finally, we present the proof of \reft{degree} based on the
application of Theorems~\ref{T:symmetry} and \ref{T:stability}.

%--------------------------------------------
\subsection{The setting}\Label{SS:setting-degree}
Let $M$ be a closed oriented manifold of odd dimension $N=2r+1$. We fix a Riemannian metric on $M$
and use it to identify the tangent and the cotangent bundles, $TM\simeq T^*M$. Let $\Lam^*TM=
\bigoplus_{j=0}^N\Lam^jTM$ denote the exterior algebra of $TM$ viewed as a vector bundle over $M$.
The space $\Ome^*(M)$ of complex-valued differential forms on $M$ coincides with the space of
sections of the complexification $\Lam^*TM\otimes\CC$ of this bundle.

The bundle $\Lam^*TM\otimes\CC$ (and, hence, the space $\Ome^*(M)$) carries 2 anti-commuting
actions of the Clifford algebra of $TM$ (the ``left'' and the ``right'' action) defined as follows
\eq{Clifford}
    c_L(v)\, \ome \ = \ v\wedge\ome-\iot_v\,\ome,
    \quad
    c_R(v)\, \ome \ = \ v\wedge\ome+\iot_v\,\ome,
    \qquad v\in TM, \ \ome\in \Ome^*(M).
\end{equation}
where $\iot_v$ denotes the interior multiplication by $v$.

The DeRham-Dirac operator $\pp$ can be written now (cf. \cite[Prop.~3.53]{BeGeVe}) as
\eq{pp}
   \pp \ = \ d+d^* \ = \
    \sum_{j=1}^N\, c_L(e_j)\, \n^{\mathrm{LC}}_{e_j},
\end{equation}
where $\n^{\mathrm{LC}}$ denotes the Levi-Civita covariant derivative and $e_1\nek e_N$ is an
orthonormal frame of $TM$.

We view the direct sum $\RR\oplus{}TM$ as a vector bundle over $M$. Consider the corresponding
sphere bundle
\eq{SS}
   \SS \ := \ \big\{\, (t,a)\in \RR\oplus TM:\, t^2+|a|^2=1\,\big\}.
\end{equation}
Let $n$ be a smooth section of the bundle $\SS$. In other words, $n=(n_0,\on)$, where $n_0\in
C^\infty(M)$, $\on\in C^\infty(M,TM)$ and $n_0^2+|n|^2=1$.

%-------
\rem{map2S}
Suppose $M$ is a parallelizable manifold, i.e., there given an identification between $TM$ and the
product $M\times\RR^N$. Then $n$ can be considered as a map
\eq{Sn}
     M \ \longrightarrow \ S^N\ := \
       \big\{\, y\in \RR^{N+1}:\, |y|^2=1\, \big\}.
\end{equation}

Also, if $M\subset\RR^{N+1}$ is a hypersurface, then, for every $x\in M$, the space
$\RR\oplus{}T_xM$ is naturally identified with $\RR^{N+1}$. Hence, $n$ again can be considered as a
map $M\to S^N$. Note, however, that, even if $M$ is parallelizable, this map is different from
\refe{Sn}.
\erem
%-------

Consider the map
\[
   \Phi:\, \RR\oplus TM \ \to \ \End \Lam^*TM\otimes\CC, \quad
   \Phi:\, n=(n_0,\on) \ \mapsto \ in_0+c_R(\on),
\]
and define the family of deformed DeRham-Dirac operators
\eq{Dm}
    D_{mn} \ = \ \pp \ + \ m\,\Phi(n):\, \Ome^*(M) \ \longrightarrow \
    \Ome^*(M).
\end{equation}

We are interested in the phase of $\Det_\tet{D_{mn}}$ for sufficiently large $m$. The following
lemma shows that this determinant is well defined.

%-------------
\lem{nokernel3}
Fix an orthonormal frame $e_1\nek e_N$ of $TM$ and set
\[
  \big|\, \n^{\mathrm{LC}}\on(x)\,\big| \ = \
     \sum_{j=1}^N\, \big|\, \n^{\mathrm{LC}}_{e_j}\on(x)\,\big|,
  \qquad
  \big|\, \n n_0(x)\,\big| \ = \
     \sum_{j=1}^N\, \big|\, \n_{e_j} n_0(x)\,\big|.
\]

For $m>\max_{x\in M}\big(|(\n^{\mathrm{LC}}\on(x)|+|\n{}n_0(x)|\big)$, zero is not in the spectrum
of $D$.
\elem
The lemma is a particular case of a more general \refl{nokernel4}, cf. below.

%---------------------------------------
\subsection{The degree of a section}\Label{SS:degree}
Note that the bundle $\SS\to M$ has a natural section $\sig:M\to \SS$, $\sig(x)=(1,0)$.

\defe{degree}
The {\em topological degree} $\deg(n)$ of the map $n$ is the
intersection number of the manifolds $\sig(M)$ and $n(M)$ inside
$\SS$.
\edefe

%---
\rem{degree}
Suppose $M$ is parallelizable and consider $n$ and $\sig$ as maps $M\to S^N$, cf. \refr{map2S}.
Then $\sig$ is the constant map $\sig(x)=(1,0)$. Hence, $\deg(n)$ is the usual topological degree
of the map $n:M\to S^N$.
\erem

%------------
\th{degree}
Let $m>\max_{x\in
  M}\big(|(\n^{\mathrm{LC}}\on(x)|+|\n{}n_0(x)|\big)$. For every
$\tet\in (0,\pi)$ such that there are no eigenvalues of $D_{mn}$ on the ray $R_\tet$ the following
equality holds
\eq{degree}
   \Det_\tet D_{mn} \ = \ (-1)^{\deg{n}}\, \big|\,\Det_\tet D_{mn}\,\big|.
\end{equation}
\eth

%---------------
\subsection{Reformulation in terms of spinors}\Label{SS:reform}
Consider the (left) {\em chirality operator}
\[
  \Gam_L \ := \ i^{\frac{N+1}2}\, c_L(e_1)\,c_L(e_2)\cdots c_L(e_N),
\]
where $e_1\nek e_N$ is an orthonormal frame of $TM$. This operator is independent of the choice of
the frame \cite[Lemma~3.17]{BeGeVe} (in fact, it coincides with the operator $\Gam$ defined in
\refe{chiral}). Moreover, $\Gam_L^2=1$ and $\Gam_L$ commutes with $c_L(v)$ and anti-commutes with
$c_R(v)$ for all $v\in TM$.

Consider the map
\[
   \widehat{}\,:\,\RR\oplus TM\ \to \ \End\Lam^*TM, \qquad
   n=(n_0,\on) \ \mapsto \hatn\ := \ \ i\,\Gam_L\, n_0 \ + \ \Gam_L\, c_R(\on).
\]
Then
\[
   \hatn^2\ =\ -\,\big(\, n_0^2+|\on|^2\,\big).
\]
Hence, the map $n\mapsto \hatn$ defines a Clifford action of $\RR\oplus{}TM$ on $\Lam^*TM$.

Assume now that $M$ is a spin-manifold (without this assumption the construction of this subsection
is true only locally, in any coordinate neighborhood). In particular, there exists a bundle
$\grS\to M$ whose fibers are isomorphic to the space of spinors over $\RR\oplus{}TM$. Then (cf.
\cite[Prop.~3.35]{BeGeVe}) there exists a bundle $\calS\to M$, such that  $\Lam^*TM\otimes\CC\to M$
can be decomposed as the tensor product $\calS\otimes\grS$, and the operators $\hatn \ (n\in
\RR\oplus{}TM)$ act only on the second factor. More precisely, if we denote by
$c_{\grS}:\RR\oplus{}TM\to \End\grS$ the Clifford action of $\RR\oplus{}TM$  on $\grS$, then
$\hatn= 1\otimes{}c_{\grS}(n)$.

We introduce now a new Clifford action $\tilc:TM\to \End\big(\Lam^*TM\otimes\CC\big)$ of $TM$ on
$\Lam^*TM\otimes\CC$, defined by the formula
\eq{tilc}
    \tilc(v) \ = \ \Gam_L\, c_L(v), \qquad v\in TM.
\end{equation}
One readily sees that $\hatn$ and $\tilc(v)$ commute for all $v\in TM, \ n\in \RR\oplus{}TM$. It
follows (cf. \cite[Prop.~3.27]{BeGeVe}) that there is a Clifford action $c_\calS:TM\to \End\calS$
such that $\tilc(v)= c_\calS(v)\otimes1$. Comparing dimensions we conclude that $\calS$ is a spinor
bundle over $M$.

It follows from \refe{pp}, that $\pp= \Gam_L\pp_{\calS}\otimes1$, where $\pp_{\calS}$ is the Dirac
operator on $\calS$. Hence, the operator \refe{Dm} takes the form
\[
  D_{mn} \ = \ \pp\ + \ m\,\Gam_L\,\hatn \ = \
  \Gam_L\,\big(\,\pp_\calS\otimes1\ + \ m\cdot 1\otimes\, c_\grS(n)\,\big):\,
   C^\infty(\calS\otimes\grS) \ \longrightarrow \ C^\infty(\calS\otimes\grS).
\]
In this form this and similar operators appeared in physical literature. In particular, for the
case when $M$ is a $(4l+1)$-dimensional sphere this operator%
\footnote{Note, however, that there is a sign discrepancy between our notation
  and the notation accepted in physical literature. Our operators
  $c_{\calS}(v), \ c_{\grS}(n)$ are skew-adjoint and satisfy the equalities
  $c_{\calS}(v)^2=-|v|^2, \ c_{\grS}(n)=-|n|^2$. Consequently, the operator $\pp_{\calS}$ is
  self-adjoint.}
was considered in \cite{AbanovWieg00}. Also a result similar to our \reft{degree} was obtained in
\cite{AbanovWieg00} for the operator
\[
    \Gam_L\cdot D_{mn} \ = \ \pp_\calS\otimes1\ + \ m\,\big(\, 1\otimes\, c_\grS(n)\,\big)
\]
on a $(4l+3)$-dimensional sphere.

%-----------------------
\subsection{The idea of the proof}\Label{SS:idea}
The rest of this section is devoted to the proof of \reft{degree}, which is based on an application
of Theorems~\ref{T:symmetry} and \ref{T:stability}. More precisely, we will deform operator
$D_{mn}$ to an operator $\tilD$ whose determinant has the same sign (in view of \reft{stability}).
We then calculate the number of imaginary eigenvalues of $\tilD$, which, in view of
\reft{symmetry}, will give us the sign of the determinants of $\tilD$ and $D_{mn}$.

First, we need to define the class of operators in which we will perform our deformation. This is
done in the next subsection.

%---------------------------
\subsection{Extension of the class of operators}\Label{SS:extension}
Let $a:M\to \RR$ and $v:M\to TM$ be a smooth function and a smooth vector field on $M$
respectively. Set
\[
   D(a,v) \ := \ \pp \ + \ ia \ + \ c_R(v).
\]
Clearly, $D_{mn}= D(mn_0,m\on)$. Also the following analogue of \refl{nokernel3} holds
%%
%---
\lem{nokernel4}
Suppose $a(x)^2 \ +\ |v(x)|^2>0$ for all $x\in M$. Fix
\[
   m_0\ >\
   \max_{x\in M}\,\big(\,|(\n^{\mathrm{LC}}v(x)|+|\n{}a(x)|\,\big).
\]
Then, for all $m\ge m_0$, zero is not in the spectrum of $D(m_0a,mv)$.
\elem
\prf
Set $\pp_m=\pp+(m-m_0)c_R(v)$. Then
\[
   D(m_0a,mv)\ =\ \pp_m \ +\ i\,m_0\, a \ +\ m_0\,c_R(v).
\]
Let $\alp\in \Ome^*(M)$. Using \refe{pp}, we obtain,
\begin{multline}\notag
  \big\|\, D(m_0a,mv)\,\alp\,\big\|^2 \\ = \ \big\|\,\pp_m\,\alp\,\big\|^2
  \ + \ m_0^2\, \|\alp\|^2 \ + \
  m_0\, \big\<\,
    \big[\, \pp_m\, (c_R(v)+ia)+(c_R(v)-ia)\,\pp_m\,\big]\,\alp,\alp
        \,\big\>
 \\ \ge \
  m_0^2\, \|\alp\|^2 \ + \ m_0\, \sum_{j=1}^N\,
   \big\<\, c_L(e_j)\,
     \big[\, c_R(\n^{\text{LC}}_{e_j}v)+
  \n_{e_j}a\,\big]\,\alp,\alp\,\big\>
  \ + \ 2\,m_0\,(m-m_0)\, \<\, |v|^2\alp,\alp\, \>
  \\ \ge \
  m_0\, \Big(\,
   m_0-\max_{x\in M}\,\big(\,|(\n^{\mathrm{LC}}v(x)|+|\n{}a(x)|\,\big)
   \,\Big)\, \|\alp\|^2.
\end{multline}
\eprf

The following lemma shows that we can apply \reft{symmetry} to the study of $\Det_\tet(a,v)$ (and,
hence, of $\Det_\tet{D_{mn}}$).

%-----------------------------
\lem{symDm}
The operators $D(a,v)$ and $-D(a,v)^*$ are conjugate to each other. Consequently, they have the
same spectral decomposition \refe{L=sumLam}.

In particular, the operators $D_{mn}$ and $-D_{mn}^*$ are conjugate to each other.
\elem
%%%%
\prf
Let $N:\Ome^*(M)\to \Ome^*(M)$ be the {\em grading operator} defined by the formula
\eq{N}
   N\,\ome \ = \ (-1)^j\,\ome, \qquad \ome\in \Ome^j(M).
\end{equation}
Then
\eq{NDN}
   N\circ D(a,v)\circ N \ = \ -\pp \ + \ ia \ - \ \,c_R(v) \ = \ -D(a,v)^*.
\end{equation}
\eprf

%-----------------------------
\subsection{Deformation of $D_{mn}$}\Label{SS:deform}
Let $n=(n_0,\on)$ be as in \reft{degree}. Suppose that $\deg(n)= \pm{k}$, where $k$ is a
non-negative integer. Then there exists a section $n'=(n_0',\on')$ of $\SS$, which is homotopic to
$n$ and has the following properties:
\begin{itemize}
\item
There exist $k$ distinct points $x_1\nek x_k\in M$ such that
\[
   n_0'(x_j)\ =\ 1,\quad \on'(x_j)\ = \ 0, \qquad j=1\nek k.
\]
\item There exists a Morse function $f:M\to \RR$ and  a
  neighborhood $U$ of the set  $\{x_1\nek x_k\}$ such that
 \[
    \on'(x) \ =\ \n{f}(x), \qquad\text{for all}\quad x\in U,
  \]
  and $\on'(x)\not=0$ for all $x\in U\backslash\{x_1\nek{}x_k\}$.
\item
If $\on'(x)=0$ and $x\not\in \{x_1\nek x_k\}$, then $n'_0(x)=-1$ and $\n{f}(x)=0$.
\end{itemize}

Let $x_{k+1}\nek x_l$ be the rest of the critical points of $f$. Then $\on'(x)\not=0$ for all
$x\not\in \{x_1\nek{}x_l\}$. Fix open neighborhoods $V_j\ (j=1\nek l)$ of $x_j$ whose closures are
mutually disjoint and such that $V_j\subset U$ for all $j=1\nek k$. We will assume that $V_j$ are
small enough so that $n'_0(x)\not=0$ and $\on'(x)\not=0$ for all $x\in V_j\backslash\{x_j\}$.

For each $j=1\nek l$ fix a neighborhood $W_j$ of $x_j$, whose closure lies inside
$V_j$.

Let $a:M\to [-1,1]$ be a smooth function such that
\eq{a(x)}
  a(x) \ = \ \begin{cases}
              \ \ 1, \quad &\text{if} \ x\in \bigcup_{j=1}^k\,W_j;\\
           -1, \quad &\text{if} \ x\not\in \bigcup_{j=1}^k\,V_j.
          \end{cases}
\end{equation}

Consider the deformation $(n_0(t),\on(t))$ of the section $(n_0',\on')\in \SS$ given by the
formulas
\begin{align}
    n_0(t) \ &= \
    \begin{cases}
    ta+(1-t)n_0', \qquad &0\le t\le1;\\
    a,\qquad    &1\le t\le 2.
    \end{cases},\notag\\
    \on(t) \ &= \
    \begin{cases}
    \on', \qquad &0\le t\le1;\\
    (t-1)\n f+(2-t)\on',\qquad    &1\le t\le 2.\notag
    \end{cases}
\end{align}
Clearly, $(n_0(t),\on(t))\not=0$ for all $t\in [0,2]$. Hence, by \refl{nokernel4}, for large $m_0$
and every $m>m_0$, $t\in [0,2]$, zero is not in the spectrum of the operator
$D\big(\,m_0n_0(t),\,m\on(t)\,\big)$. \reft{stability} implies now that the determinant of $D_{mn}$
has the same sign as the determinant of
\[
    D\big(\,m_0n_0(2),\,m\on(2)\,\big) \  =  \ D(m_0a,m\n{f}).
\]

%-------------------------------------------------
\subsection{The spectrum of the operator $D(0,m\n{f})$}
\Label{SS:D0nf}
Before investigating the operator  \linebreak $D(m_0a,m\n{f})$ we consider a simpler operator
\[
   D(0,m\n{f}) \ = \ \pp \ + \ m\,c_R(\n f) \ = \
    e^{-mf}\, d\ e^{mf} \ + \ \big(\,  e^{-mf}\, d\ e^{mf}\,\big)^*.
\]
This is a self-adjoint operator whose spectrum was studied by Witten \cite{Witten82} (see, for
example, \cite{Shubin96Morse} for a mathematically rigorous exposition of the subject).  In
particular, $D(0,m\n{f})$ has the following properties:
\begin{itemize}
\item There exist a constant $C>0$ and a function $r(m)>0$ such that
  $\lim_{t\to0}r(m)= 0$ and, for all sufficiently large $m>0$, the
  spectrum of $D(0,m\n{f})$ lies inside the set
  \[
     \big(-\infty,\, -C\sqrt{m}\,\big)\, \cup\, \big(\,-r(m),r(m)\,\big)\, \cup\,
       \big(\,C\sqrt{m},\, \infty).
  \]
\item Let $E_m$ denote the span of the eigenvectors of
  $D(0,m\n{f})$
  with eigenvalues in the interval ${\big(-r(m),r(m)\big)}$. Then, for all sufficiently large $m$, the space $E_m$
  has a basis $\alp_{1,m}\nek\alp_{l,m}$ ($\|\alp_{j,m}\|= 1$) such that
  each $\alp_{j,m}$ {\em is concentrated in $W_j$} in the following sense:
  \eq{concentr}
     \int_{W_j}\, \alp_{j,m}\wedge*\alp_{j,m} \ = \ 1 \ - \ o(1),
     \qquad \text{as}\quad m\to\infty.
  \end{equation}
  (Here $o(1)$ stands for a vector whose norm tends to 0 as
  $m\to\infty$).
  In particular, $\dim{E_m}=l$.
\end{itemize}
Note, that \refe{concentr} and \refe{a(x)} imply that
\eq{concentr2}
   a(x)\, \alp_{j,m}(x) \ = \
     \begin{cases}
       \alp_{j,m}(x)+o(1), \quad&\text{for}\ j=1\nek k;\\
       -\alp_{j,m}(x)+o(1), \quad&\text{for}\ j=k+1\nek l.
     \end{cases}
\end{equation}

%-------------------------------------------------
\subsection{The spectrum of the operator $D(m_0a,m\n{f})$}\Label{SS:Dmm0}
Let $m_0$ be as in \refss{deform} and let $C$ be as in \refss{D0nf}. Choose $m$ large enough so
that
\[
  m \ > \ \left(\, \frac{4(m_0+1)}C\,\right)^2.
\]
and $r(m)<1$. We view the operator
\[
  D(m_0a,m\n{f}) \ = \ D(0,m\n{f}) \ + \ i\,m_0\, a(x)
\]
as a perturbation of $D(0,m\n{f})$.

%--------------------------
\lem{spindisk}
The number of eigenvalues $\lam$ (counting with multiplicities) of $D(m_0a,m\n{f})$ which satisfy
\[
  |\lam| \ < \ 2(m_0+1), \quad \IM\lam\ > \ 0,
\]
is equal to $k=\deg{n}$.
\elem
%%%%
\prf
The spectral projection of the operator $D(0,m\n{f})$ onto the space $E_m$ (cf. \refss{D0nf}) is
given by the Cauchy integral
\eq{Pm}
   P_m \ = \ \frac{1}{2\pi i}\,
    \oint_\gam\, \big(\, \lam-D(0,m\n{f})\,\big)^{-1}\, d\lam,
\end{equation}
where $\gam$ is the boundary of the disk $B=\big\{\, z\in\CC:\, |z|<2(m_0+1)\,\big\}$.

Note that, for all $\lam\in \gam$, we have
\eq{normresolv}
   \left\|\,\big(\, \lam-D(0,m\n{f})\,\big)^{-1}\, \right\| \ \le \
   \frac1{\dist\big(\,\lam,\spec D(0,m\n{f})\,\big)} \ = \
   \frac1{2m_0+1}.
\end{equation}
Hence, for all $\lam \in \gam$, we obtain
\begin{multline}\Label{E:normresolv2}
   \left\|\,\big(\, \lam-D(m_0a,m\n{f})\,\big)^{-1}\, \right\| \\ \le \
   \left\|\,\big(\, \lam-D(0,m\n{f})\,\big)^{-1}\, \right\|\cdot
   \big\|\,\big(\,
      1-(\lam-D(0,m\n{f}))^{-1}m_0a\,\big)^{-1}\, \big\| \\ \le \
   \frac1{2m_0+1}\cdot\frac1{1-\frac{m_0}{2m_0+1}}
   \ = \ \frac1{m_0+1}.
\end{multline}
In particular, $\gam$  is contained in the resolvent set of $D(m_0a,m\n{f})$.

Let $E_m'$ denote the span of the root vectors of $D(m_0a,m\n{f})$ with eigenvalues in $B$. The
spectral projection of $D(m_0a,m\n{f})$ onto $E_m'$ is given by the formula
\[
   P_m' \ = \ \frac{1}{2\pi i}\,
    \oint_\gam\, \big(\, \lam-D(m_0a,m\n{f})\,\big)^{-1}\, d\lam.
\]
Using \refe{normresolv} and \refe{normresolv2}, we obtain
\begin{multline}\Label{E:P-P'}
    \big\|\, P_m-P_m'\,\big\| \ = \
    \frac{1}{2\pi}\, \left\|\, \oint_\gam\,
    \big(\, \lam-D(0,m\n{f})\,\big)^{-1}\, m_0a\,
                 \big(\, \lam-D(m_0a,m\n{f})\,\big)^{-1}\,d\lam\,
    \right\|
    \\ \le \
    2(m_0+1)\cdot \frac1{2m_0+1}\cdot m_0\cdot\frac1{m_0+1}
    \ = \ \frac{2m_0}{2m_0+1} \ < \ 1.
\end{multline}
In particular,
\[
   \dim E_m' \ = \ \dim E_m \ = \ l,
\]
and the projection $P_m'$ maps $E_m$ isomorphically onto $E_m'$. Recall that the basis
$\alp_{1,m}\nek \alp_{l,m}$ of $E_m$ was defined in \refss{D0nf}. Then \/ $P'_m\alp_{1,m}\nek
P_m'\alp_{l,m}$ is a basis of $E_m'$.

From \refe{concentr2}, we get
\[
  D(m_0a,m\n{f})\, \alp_{j,m} \ = \
   \begin{cases}
     i\,m_0\,\alp_{j,m} + o(1),\quad&\text{for}\ j=1\nek k;\\
     -i\,m_0\,\alp_{j,m} + o(1),\quad&\text{for}\ j=k+1\nek l.
   \end{cases}
\]
Since the operators $D(m_0a,m\n{f})$ and $P_m'$ commute we obtain
\[
  D(m_0a,m\n{f})\, P_m'\alp_{j,m} \ = \
   \begin{cases}
     i\,m_0\,P_m'\alp_{j,m} + o(1),\quad&\text{for}\ j=1\nek k;\\
     -i\,m_0\,P_m'\alp_{j,m} + o(1),\quad&\text{for}\ j=k+1\nek l.
   \end{cases}
\]
Hence, the restriction of $D(m_0a,m\n{f})$ to $E_m'$ has exactly $k$ eigenvalues (counting with
multiplicities) with positive imaginary part.
\eprf

%------------------------------------
\subsection{Proof of \reft{degree}}\Label{SS:prdegree}
Clearly, all the eigenvalues of $D(m_0a,m\n{f})$ satisfy
\eq{IMlam}
  \big|\, \IM \lam\,\big| \ \le \ m_0.
\end{equation}
In particular, all the eigenvalues of $D(m_0a,m\n{f})$  which lie on the ray $R_{\pi/2}$ belong to
the disc $B=\big\{\, z\in\CC:\, |z|<2(m_0+1)\,\big\}$.  Since the spectrum of $D(m_0a,m\n{f})$ is
symmetric with respect to the imaginary axis the number of these eigenvalues (counting with
multiplicities) has the same parity as the number of all eigenvalues, which lie in $B$ and have
positive imaginary part. \reft{degree} follows now from \reft{symmetry} and \refl{spindisk}.
\hfill$\square$

%---------------------------------------------------------
%---------------------------------------------------------

%\nocite{*}
\providecommand{\bysame}{\leavevmode\hbox to3em{\hrulefill}\thinspace}
\providecommand{\MR}{\relax\ifhmode\unskip\space\fi MR }
\providecommand{\MRhref}[2]{%
  \href{http://www.ams.org/mathscinet-getitem?mr=#1}{#2}
} \providecommand{\href}[2]{#2}


\begin{thebibliography}{10}

\bibitem{Abanov-Hopf}
A.~G. Abanov, \emph{Hopf term induced by fermions}, Phys.Lett. \textbf{B492}
  (2000), 321--323.

\bibitem{AbanovWieg00}
A.~G. Abanov and P.~B. Wiegmann, \emph{Theta-terms in nonlinear sigma-models},
  Nucl.Phys. \textbf{B570} (2000), 685--698.

\bibitem{Agranovich90}
M.~S. Agranovich, \emph{Elliptic operators on closed manifolds}, Current
  problems in mathematics. Fundamental directions, Vol.\ 63 (Russian), Itogi
  Nauki i Tekhniki, Akad. Nauk SSSR Vsesoyuz. Inst. Nauchn. i Tekhn. Inform.,
  Moscow, 1990, pp.~5--129.

\bibitem{BeGeVe}
N.~Berline, E.~Getzler, and M.~Vergne, \emph{Heat kernels and {Dirac}
  operators}, Springer-Verlag, 1992.


\bibitem{Berry84}
M. Berry, {\em Quantal phase factors accompanying adiabatic changes},
Proc. R. Soc. Lond.  A {\bf  392}, (1984), 45-57.


\bibitem{BFK91}
D.~Burghelea, L.~Friedlander, and T.~Kappeler, \emph{On the determinant of
  elliptic differential and finite difference operators in vector bundles over
  {$S\sp 1$}}, Comm. Math. Phys. \textbf{138} (1991), no.~1, 1--18.

%\bibitem{Guillemin85}
%V.~Guillemin, \emph{A new proof of {W}eyl's formula on the asymptotic
%  distribution of eigenvalues}, Adv. in Math. \textbf{55} (1985).

\bibitem{Kato}
T.~Kato, \emph{Perturbation theory for linear operators}, Springer-Verlag,
  1966.

\bibitem{Lidskii59}
V.~B. Lidski\u{\i}, \emph{Non-selfadjoint operators with a trace}, Dokl. Akad.
  Nauk SSSR \textbf{125} (1959), 485--487.

\bibitem{Markus88}
A.~S. Markus, \emph{Introduction to the spectral theory of polynomial operator
  pencils}, Translations of Mathematical Monographs, vol.~71.

\bibitem{Ponge-asymetry}
R.~Ponge, \emph{Spectral asymetry, zeta function and the noncommutative
  residue}, Preprint.


\bibitem{Redlich84}
A. N. Redlich, {\em Gauge Noninvariance and Parity Nonconservation of
Three-Dimensional Fermions}, Phys. Rev. Lett. {\bf 52}, (1984),
18-21.

\bibitem{Redlich84a}
\bysame, {\em Parity violation and gauge noninvariance of the
effective gauge field action in three dimensions}, Phys. Rev. D {\bf
29}, (1984), 2366-2374.

\bibitem{Retherford93}
J.~R. Retherford, \emph{Hilbert space: compact operators and the trace
  theorem}, London Mathematical Society Student Texts, vol.~27, Cambridge
  University Press, Cambridge, 1993.

\bibitem{Seeley67}
R.~Seeley, \emph{Complex powers of elliptic operators}, Proc. Symp. Pure and
  Appl. Math. AMS \textbf{10} (1967), 288--307.

\bibitem{ShubinPDObook}
M.~A. Shubin, \emph{Pseudodifferential operators and spectral theory}, Springer
  Verlag, Berlin, New York, 1980.

\bibitem{Shubin96Morse}
\bysame, \emph{Semiclassical asymptotics on covering manifolds and {M}orse
  inequalities}, Geom. Funct. Anal. \textbf{6} (1996), 370--409.

\bibitem{Witten82}
E.~Witten, \emph{Supersymmetry and {Morse} theory}, J. of Diff. Geom.
  \textbf{17} (1982), 661--692.

\bibitem{Wodzicki84}
M.~Wodzicki, \emph{Local invariants of spectral asymmetry}, Invent. Math.
  \textbf{75} (1984), no.~1, 143--177.

\bibitem{Wodzicki87}
\bysame, \emph{Noncommutative residue. {I}. {F}undamentals}, $K$-theory,
  arithmetic and geometry (Moscow, 1984--1986), Lecture Notes in Math., vol.
  1289, Springer, Berlin, 1987, pp.~320--399.

\bibitem{Wojciechowski99}
K.~P. Wojciechowski, \emph{Heat equation and spectral geometry. {I}ntroduction
  for beginners}, Geometric methods for quantum field theory (Villa de Leyva,
  1999), World Sci. Publishing, River Edge, NJ, 2001, pp.~238--292.

\end{thebibliography}
\end{document}